\documentclass[aps,prl,preprint,superscriptaddress]{revtex4-1}
\usepackage{graphicx,subfigure,amsmath,bbm}
\usepackage{dcolumn}
\usepackage{bm}
\usepackage{amsmath}
\usepackage{hyperref}
\usepackage{amsmath}
\usepackage{natbib}
\begin{document}

\title{Ghost Imaging at an XUV Free-Electron Laser}

\author{Young Yong Kim}
\affiliation{Deutsches Elektronen-Synchrotron (DESY), Notkestra\ss{e} 85, D-22607 Hamburg, Germany}
\author{Luca Gelisio}
\affiliation{Deutsches Elektronen-Synchrotron (DESY), Notkestra\ss{e} 85, D-22607 Hamburg, Germany}
\author{Giuseppe Mercurio}
\affiliation{Department Physik, Universit\"a{t} Hamburg, Luruper Chaussee 149, D-22761 Hamburg, Germany}
\affiliation{Center for Free-Electron Laser Science, Luruper Chaussee 149, D-22761 Hamburg, Germany}
\affiliation{Present address: European XFEL GmbH, Holzkoppel 4, D-22869 Schenefeld, Germany}
\author{Siarhei Dziarzhytski}
\affiliation{Deutsches Elektronen-Synchrotron (DESY), Notkestra\ss{e} 85, D-22607 Hamburg, Germany}
\author{Martin Beye}
\affiliation{Deutsches Elektronen-Synchrotron (DESY), Notkestra\ss{e} 85, D-22607 Hamburg, Germany}
\author{Lars Bocklage}
\affiliation{Deutsches Elektronen-Synchrotron (DESY), Notkestra\ss{e} 85, D-22607 Hamburg, Germany}
\affiliation{The Hamburg Centre for Ultrafast Imaging, Luruper Chaussee 149, D-22761 Hamburg, Germany}
\author{Anton Classen}
\affiliation{Institut f\"u{r} Optik, Information und Photonik, Universit\"a{t} Erlangen-N\"u{r}nberg, Staudtstra\ss{e} 1, D-91058 Erlangen, Germany}
\affiliation{Erlangen Graduate School in Advanced Optical Technologies (SAOT), Universit\"a{t} Erlangen-N\"u{r}nberg, Paul-Gordan-Staudtstra\ss{e} 6, D-91052 Erlangen, Germany}
\author{Christian David}
\affiliation{Paul Scherrer Institut, Villigen PSI, 5232, Switzerland}
\author{Oleg Yu. Gorobtsov}
\affiliation{Deutsches Elektronen-Synchrotron (DESY), Notkestra\ss{e} 85, D-22607 Hamburg, Germany}
\affiliation{Present address: Department of Materials Science and Engineering, Cornell University, Ithaca, NY 14850, USA}
\author{Ruslan Khubbutdinov}
\affiliation{Deutsches Elektronen-Synchrotron (DESY), Notkestra\ss{e} 85, D-22607 Hamburg, Germany}
\affiliation{National Research Nuclear University MEPhI (Moscow Engineering Physics Institute), Kashirskoe shosse 31, 115409 Moscow, Russia}
\author{Sergey Lazarev}
\affiliation{Deutsches Elektronen-Synchrotron (DESY), Notkestra\ss{e} 85, D-22607 Hamburg, Germany}
\affiliation{National Research Tomsk Polytechnic University (TPU), pr. Lenina 30, 634050 Tomsk, Russia}
\author{Nastasia Mukharamova}
\affiliation{Deutsches Elektronen-Synchrotron (DESY), Notkestra\ss{e} 85, D-22607 Hamburg, Germany}
\author{Yury N. Obukhov}
\affiliation{Theoretical Physics Laboratory, Nuclear Safety Institute, Russian Academy of Sciences, B. Tulskaya 52, 115191 Moscow, Russia}
\author{Benedikt R\"osner}
\affiliation{Paul Scherrer Institut, Villigen PSI, 5232, Switzerland}
\author{Kai Schlage}
\affiliation{Deutsches Elektronen-Synchrotron (DESY), Notkestra\ss{e} 85, D-22607 Hamburg, Germany}
\author{Ivan A. Zaluzhnyy}
\affiliation{Deutsches Elektronen-Synchrotron (DESY), Notkestra\ss{e} 85, D-22607 Hamburg, Germany}
\affiliation{Present address: University of California, San Diego, La Jolla, California 92093, USA}
\author{G\"unter Brenner}
\affiliation{Deutsches Elektronen-Synchrotron (DESY), Notkestra\ss{e} 85, D-22607 Hamburg, Germany}
\author{Ralf R\"ohlsberger}
\affiliation{Deutsches Elektronen-Synchrotron (DESY), Notkestra\ss{e} 85, D-22607 Hamburg, Germany}
\affiliation{The Hamburg Centre for Ultrafast Imaging, Luruper Chaussee 149, D-22761 Hamburg, Germany}
\author{Joachim von Zanthier}
\affiliation{Institut f\"u{r} Optik, Information und Photonik, Universit\"a{t} Erlangen-N\"u{r}nberg, Staudtstra\ss{e} 1, D-91058 Erlangen, Germany}
\affiliation{Erlangen Graduate School in Advanced Optical Technologies (SAOT), Universit\"a{t} Erlangen-N\"u{r}nberg, Paul-Gordan-Staudtstra\ss{e} 6, D-91052 Erlangen, Germany}
\author{Wilfried Wurth}
\affiliation{Deutsches Elektronen-Synchrotron (DESY), Notkestra\ss{e} 85, D-22607 Hamburg, Germany}
\affiliation{Department Physik, Universit\"a{t} Hamburg, Luruper Chaussee 149, D-22761 Hamburg, Germany}
\affiliation{Center for Free-Electron Laser Science, Luruper Chaussee 149, D-22761 Hamburg, Germany}
\author{Ivan A. Vartanyants}
\email{Corresponding author: ivan.vartaniants@desy.de}
\affiliation{Deutsches Elektronen-Synchrotron (DESY), Notkestra\ss{e} 85, D-22607 Hamburg, Germany}
\affiliation{National Research Nuclear University MEPhI (Moscow Engineering Physics Institute), Kashirskoe shosse 31, 115409 Moscow, Russia}

\date{\today}

\begin{abstract}
Radiation damage is one of the most severe resolution limiting factors in x-ray imaging, especially relevant to biological samples.
One way of circumventing this problem is to exploit correlation-based methods developed in quantum imaging.
Among these, there is ghost imaging (GI) in which the image is formed by radiation that has never interacted with the sample.
Here, we demonstrate GI at an XUV free-electron laser by utilizing correlation techniques.
We discuss the experimental challenges, optimal setup, and crucial ingredients to maximize the achievable resolution.
\end{abstract}


\maketitle

\newpage

In recent years x-ray microscopy has witnessed dramatic progress both in the scanning and full-field mode due to the availability of high brightness 3-rd generation synchrotron sources in combination with improved x-ray focusing optics~\cite{Sakdinawat2010_XMicro, Burkhard2011_XMicro}.
Spectacular improvement in x-ray microscopy were reported in particular in material science and biology using new imaging schemes adopted from visible microscopy~\cite{Holt2013, Larabell2010}.
However, certain limitations in resolution as well as radiation damage, especially affecting biological samples, inherent to these techniques triggered the search for alternative methods.
In particular, among them are lens-less approaches, exploiting the coherence properties of the x-ray beams produced by modern synchrotron sources~\cite{Miao1999, Robinson2001, Chapman2010, seeck2015x}.
Careful analysis of radiation damage produced in protein crystallography and imaging of biological samples recently demonstrated that an ultimate limit of $10$ nm in resolution can be potentially achieved with these sources~\cite{Howells2009}.
The advent of x-ray free-electron lasers (XFELs) allowed to establish even more advanced concepts of imaging biological samples, e.g., employing the method of "diffraction-before-destruction", where diffraction patterns are obtained before the samples are disintegrated due to Coulomb explosion~\cite{Neutze2000, Chapman2006}.
This concept looks especially attractive for imaging biological samples, giving birth to single particle imaging (SPI) experiments at XFELs~\cite{Neutze2000}.
Yet, due to various experimental constraints, progress in SPI was slower than expected~\cite{Aquila2015}, with a resolution below 10 nm in imaging of 70 nm viruses only recently reported~\cite{Max_Iucrj2018}.
This is still far beyond the ultimate goal of obtaining atomic resolution of macromolecules.
The present situation thus calls for the exploration of non-traditional approaches to x-ray imaging employing XFELs.

With the advent of the laser in the 1960s the new domain of quantum optics emerged~\cite{Glauber1963}, which became one of the most dynamic fields of optical science using visible light~\cite{Agarwal_QO2013}.
Within quantum optics the subject of quantum imaging, based on utilizing correlation techniques, is presently evolving  particularly rapidly~\cite{Belinskii1994, Sergienko1995, Malvin1997, Malvin2001, Robert2002, Lugiato2004, Shi-Yao2005, Shih2005, Zhang2005, Thiel_PRL2007, Erkmen2010, Oppel_PRL2012, Classen_PRL2016, Pelliccia2016, Yu2016, Schneider2017, Pelliccia2018, Zhang_Optica2018, Li_PRL2018} [see for review~\cite{Genovese2016}].
Employing techniques of quantum imaging may lead to an enhanced resolution that can, in principle, overcome Abbe's fundamental limit of classical optics~\cite{Thiel_PRL2007, Oppel_PRL2012, Classen_PRL2016, Schneider2017}.
Interestingly, it was recently demonstrated that, in spite of its name, quantum imaging protocols can be implemented also by use of classical light~\cite{Robert2002, Shih2005, Thiel_PRL2007, Oppel_PRL2012, Classen_PRL2016}, allowing for the extension of the protocols to shorter wavelengths using synchrotron and, in particular, FEL sources~\cite{Schneider2017}.

\begin{figure}[h!]
	
	\centering
	\includegraphics[scale = 0.4]{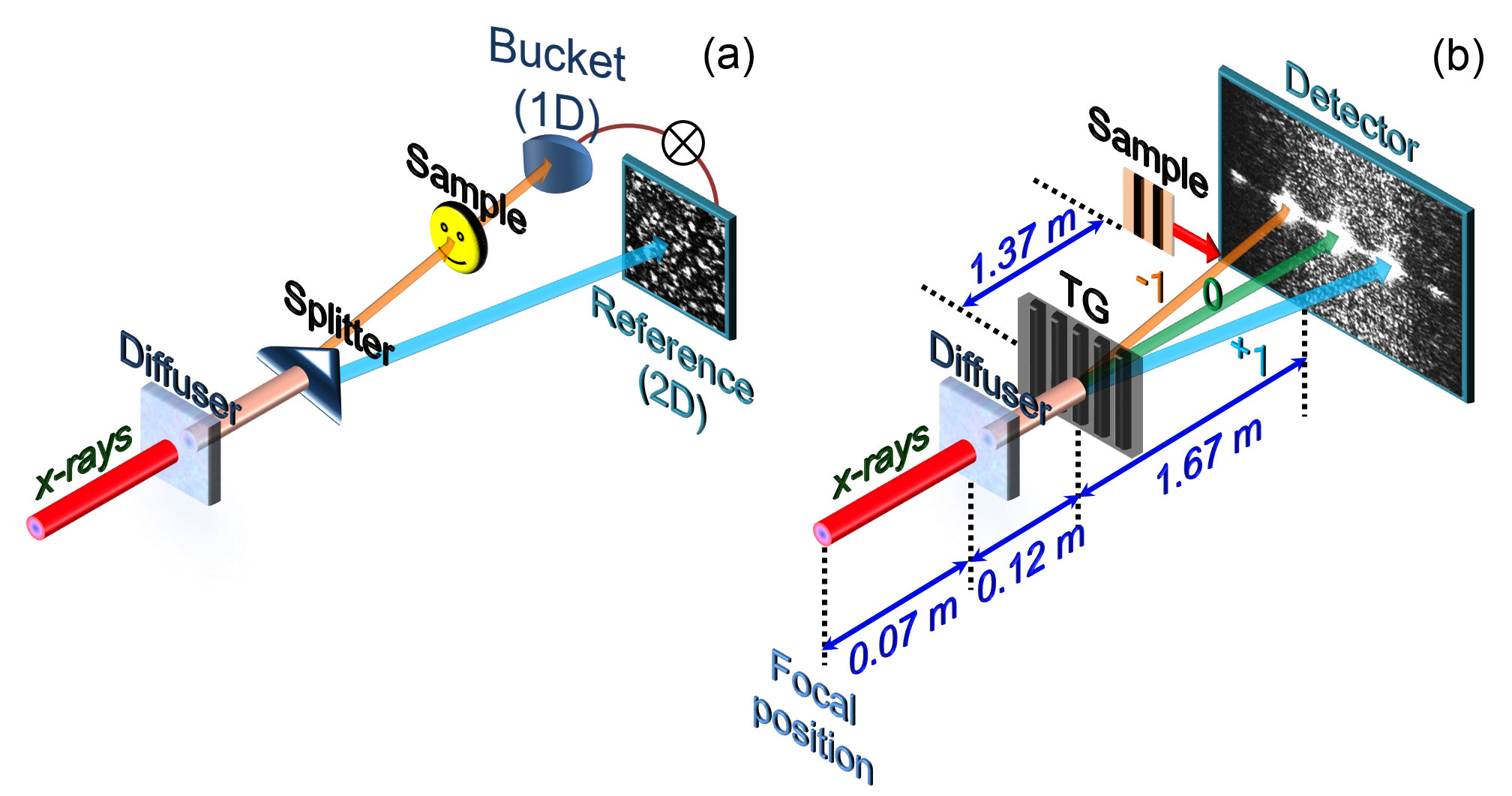}
	\caption{Conceptual layout of a ghost imaging experiment (a) and schematic setup used at FLASH, which allowed successful ghost imaging using an FEL beam (b).
	}
	\label{fig:GI}
\end{figure}

One of the schemes proposed and developed in the domain of quantum imaging is the so-called ghost imaging (GI)~\cite{Sergienko1995, Robert2002} [see for review~\cite{Erkmen2010}].
The idea is based on parallel measurement of two correlated beams [see Fig.~\ref{fig:GI}(a)], where one beam passes through the object and the transmitted light is detected using a bucket detector, whereas the other beam is freely propagating and recorded using a pixelized detector.
Coincidence measurements are then performed between each pixel of the pixelized and bucket detectors, so that the object is finally reconstructed by coincident detection of the two photons at both detectors.
Initially it was assumed that GI needs quantum correlations between the two beams~\cite{Sergienko1995}, but later it was realized that measurements can also be performed using classically correlated coherent light beams~\cite{Robert2002}, pseudothermal light~\cite{Lugiato2004, Shi-Yao2005, Shih2005}, or even thermal sources~\cite{Zhang2005}.
Recently, it has been demonstrated that GI can also be implemented using hard x-ray radiation from synchrotron sources~\cite{Pelliccia2016, Yu2016, Pelliccia2018}.
One of the appealing features of GI is that, by varying the ratio of the photon flux in the two arms, one can substantially lower the radiation dose absorbed by the sample while increasing the intensity of the other beam that does not impinge on the sample.
As such, radiation damage, that is the most severe factor limiting the resolution in biological imaging, may be alleviated.

GI with thermal light sources is based on correlating speckles, i.e. intensity fluctuations of the incident beam, which can be produced either by illuminating with coherent light a diffuser \cite{Martienssen_AJP1964, Estes_JOSA1971}, or utilizing the natural fluctuations of thermal light, as in the original experiment by Hanbury Brown and Twiss (HBT)~\cite{HBT1956, HBT1956_2}.
The correlation of the intensity in the two arms forms the ghost image of the object $T_{GI}$ according to~\cite{Katz2009}
\begin{equation}
\label{eq:GI}
T_{GI}(x,y) = \langle(B - \langle B \rangle_M)I_{r}(x,y)\rangle_M \ ,
\end{equation}
where $B$ is the bucket signal [the total intensity transmitted through the sample], $I_{r}(x,y)$ is the intensity collected at each pixel of the reference detector, and  $\langle...\rangle_M$ expresses the average value over $M$ realizations.

GI experiments using synchrotron radiation are challenging mostly due to the limited photon flux per synchrotron pulse which strongly degrades the quality of the images~\cite{Pelliccia2016}.
On the other hand, performing such an experiment at a FEL has many striking advantages.
First, coincident detection of the two images in the two arms of the splitted beam is inherently provided by the short [femtosecond] pulsed operation of the FEL radiation.
Second, it is comparably easy to obtain Fourier limited pulses from FELs that provide 100\% contrast in HBT interferometry~\cite{Singer_PRL2013, Oleg_PRA2017, Oleg_Nat_Comm_2018}; the latter should lead to strongly enhanced contrast in the GI experiments.
Third and most importantly, the Fourier limited pulses produced at FELs have very high intensities [e.g. $10^{9}$ photons per pulse at FLASH~\cite{Singer_PRL2013}], i.e., many orders of magnitude more than that can be delivered by synchrotron sources.
Here, we report on the realization of a GI experiment performed at an extreme ultraviolet (XUV) FEL source.

For our GI experiments we employed the PG2 beamline of the FLASH facility~\cite{FLASH1}, using the single bunch mode with 10 Hz repetition rate.
The beam from the undulator with average pulse energy of 23 $\mu$J went through a Xe gas absorber at $3{\times}10^{-2}$ mbar reducing the pulse power by about two orders of magnitude and thus avoiding radiation damage of the detector.
The transmitted [0th order] beam from the monochromator at a wavelength of 13.4 nm [92.5 eV photon energy] had a bandwidth of $\sim$1\% and a footprint of approximately $50{\times}50$ $\mu \mathrm{m}^{2}$ full width at half maximum (FWHM) at the focal position, 1.86 m in front of the detection position~\cite{PG2beamline, PG2beamline2}.
The end-station, consisting of a diffuser, transmission grating (TG), sample, and detector, is schematically shown in Fig.~\ref{fig:GI}(b).
As mentioned, GI at classical sources relies on a time-varying speckle pattern, used to probe the sample.
In our experiment, the speckles were generated by coherently illuminating a diffuser made of silica nanospheres of approximately 400 nm in diameter~\cite{Schneider2017}.
The diffuser was continuously moved such that each FEL pulse was impinging on a different transverse position of the diffuser so that the speckle pattern effectively varied from pulse to pulse.
As sample we employed a 300 nm thick cobalt two-slit structure with slit bars of 200 $\mu$m width and 1.5 mm height separated by 200 $\mu$m.
The structure was prepared on 100 nm thick Si$_3$N$_4$ membranes by laser lithography, sputter deposition and lift-off processing.
The relative transmission of the cobalt film was about $10^{-8}$ at the wavelength of 13.4 nm.
The sample was installed 1.37 m downstream of the TG and was illuminated by the grating's first order.

The intensities of the bucket and reference signal, at the end of the experimental unit, were measured simultaneously by an in-vacuum Andor Ikon charge-coupled device (CCD) composed of $2048{\times}2048$ pixels with $13.5{\times}13.5$ $\mu$m$^2$ size each, positioned 0.30 m downstream from the sample.
The detector was operated at a repetition rate of 5 Hz and triggered by the PG2 beamline fast shutter.
The bucket and reference signals were defined as regions of $80{\times}80$ pixels on the CCD.
The intensity over the area corresponding to the bucket detector was then integrated to mimic the signal of a large area bucket detector.
The direct beam [0th order from the TG] was blocked by the sample holder in order to avoid detector saturation.

\begin{figure}[h!]
	
	\centering
	\includegraphics[scale = 0.58]{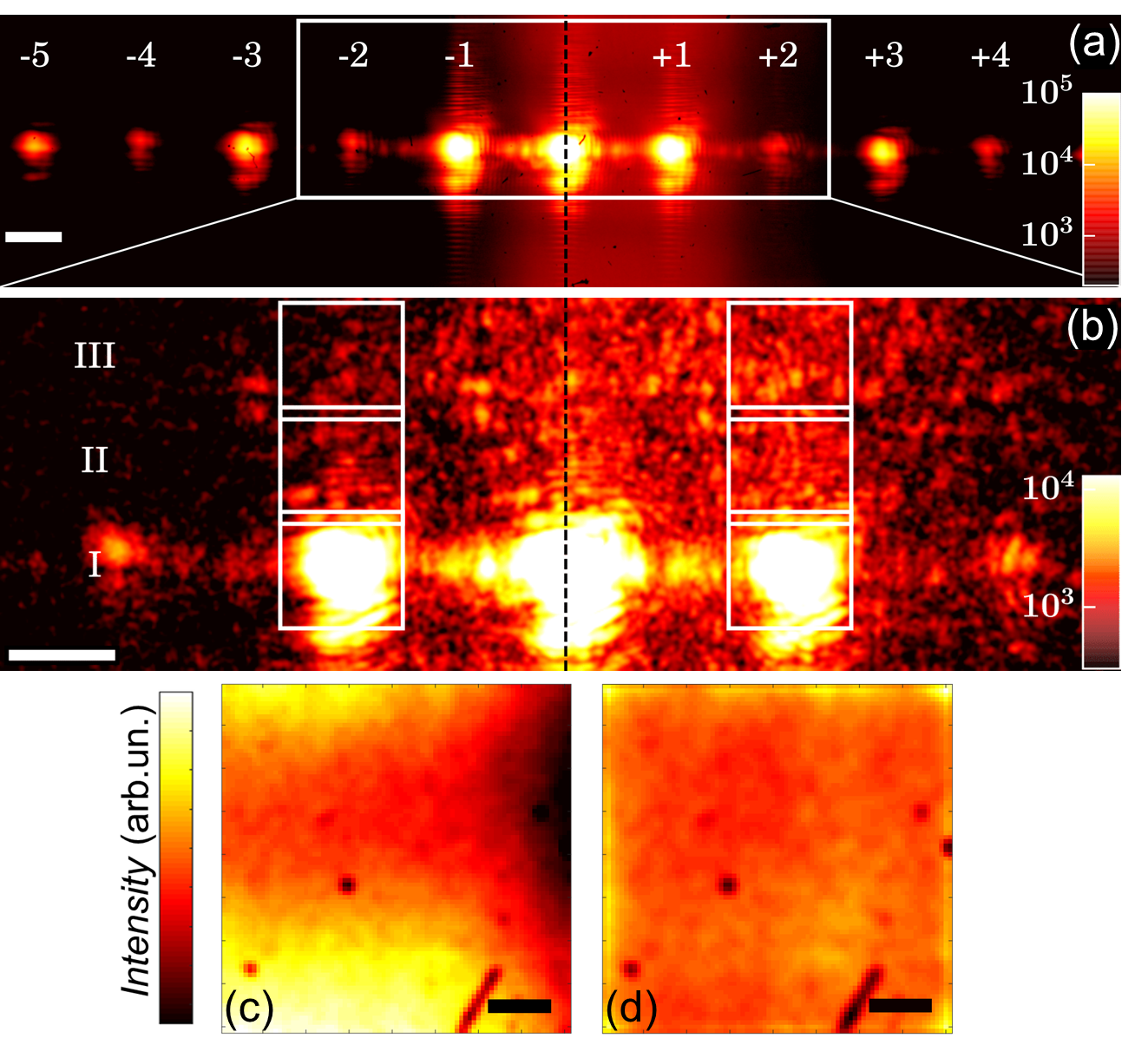}
	\caption{
		(a) Time-averaged intensity with 2,115 realizations.
		(b) Single pulse speckle pattern obtained at a selected position of the diffuser in the region shown by white rectangle in (a).
		White boxes define different probed positions.
		In [(a), (b)] the colorbar defines the number of detected photons and the length of the white bar is equal to one millimeter.
		[(c), (d)] Time-averaged intensity [20,000 realizations] detected in position III before (c) and after (d) the intensity normalization [see text for details].
		In [(c), (d)] dark clusters of pixels are detector defects and the black bar is equal to 200~$\mu$m.
	}
	\label{fig:beam}
\end{figure}

The most important ingredient for a successful implementation of GI is that the speckle patterns of both arms of the beam are identical and non-overlapping [see Fig.~\ref{fig:GI}(a)].
To reach this goal, a TG made of a $350$ nm free-standing gold structure with a pitch size of $17$ $\mu$m and a slit width of $10.2$ $\mu$m was placed in the beam to produce identical copies of the incoming light.
An example of the measured intensity distribution behind the TG, averaged over 2,115 realizations in absence of the sample, is shown in Fig.~\ref{fig:beam}(a).
Diffraction up to the 5th order is well visible, with the beam size (FWHM) at the first orders being about $340$ $\mu$m.
Three additional features can be seen.
First, a stripy structure of the beam extended in the vertical direction, due to specific TG enforcement technology, can be observed.
Second, a partial overlap of intensities from different diffraction orders is visible.
Third, a small asymmetry in the intensity distribution with respect to the grating symmetry plane of the 0th order [black dashed line in Fig.~\ref{fig:beam}(a,b)] was detected.
While the latter was possibly caused by a misalignment of the TG, the others resulted from the fact that the TG was not specifically designed for the wavelength of the experiment.

A typical speckle pattern produced by a single FEL pulse after the diffuser is shown in Fig.~\ref{fig:beam}(b).
The speckle size was on the order of 100~$\mu$m FWHM.
Positions at different vertical distances from the diffraction plane, defined in Fig.~\ref{fig:beam}(b) as regions I, II, and III, were probed to investigate the effect of the direct beam on the GI reconstruction.
It turned out that we were not able to reconstruct any ghost image at positions I and II due to the presence of a strong background, especially evident when considering the average of intensities over all realizations [Fig.~\ref{fig:beam}(a) and Appendix A].
On the contrary, position III was less affected by the direct beam, and mostly consisted of speckles on top of a background which was the most homogeneous among the three cases [see Fig.~\ref{fig:beam}(c)].

\begin{figure}[h!]
	
	\centering
	\includegraphics[scale = 0.45]{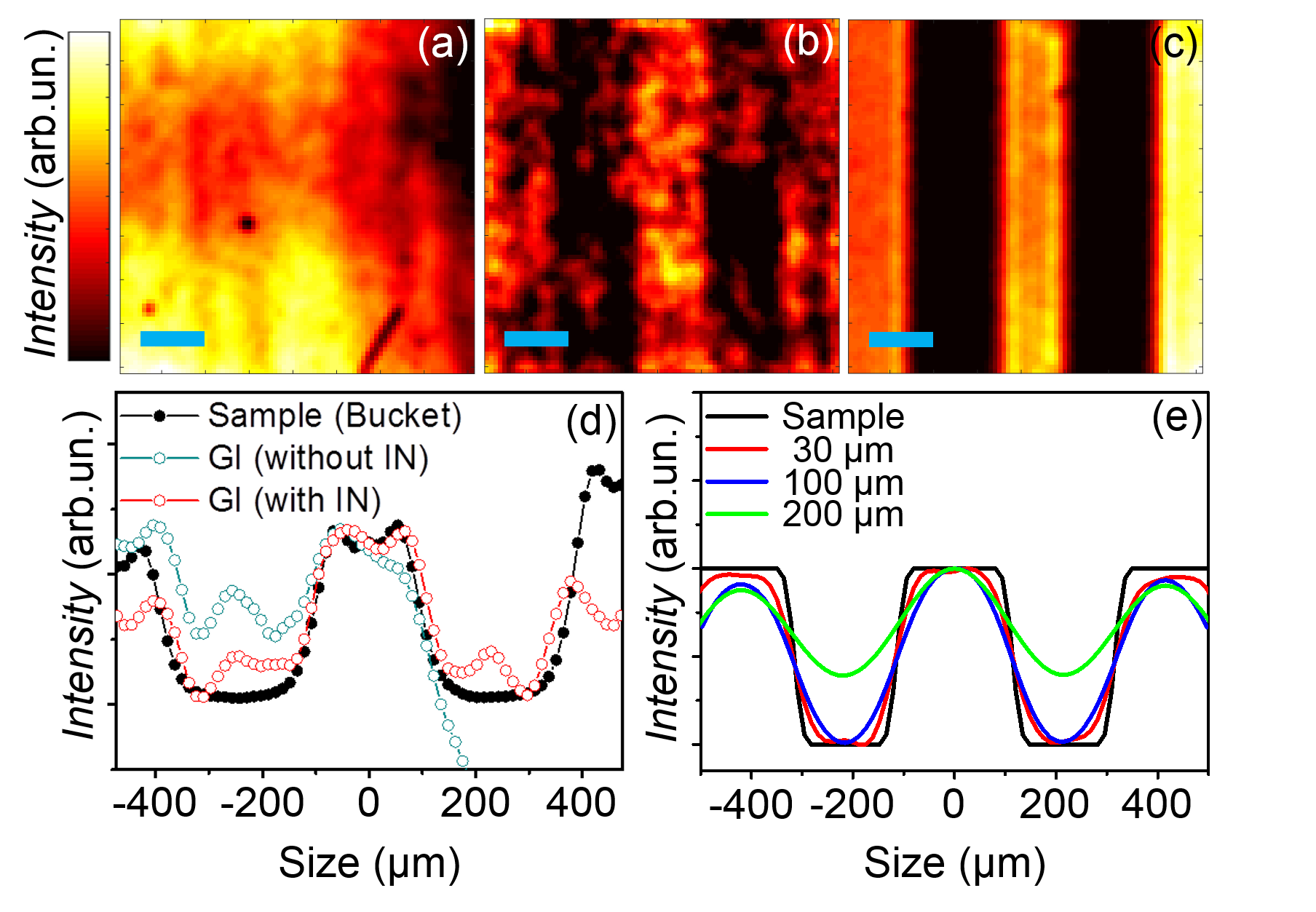}
	\caption{[(a), (b)] Results of ghost imaging reconstruction without intensity normalization of reference signal (a) and the same result with the intensity normalization of the reference signal (b).
		(c) Time-average image of 20,000 frames at the bucket detector position.
		Intensities in panels [(a)-(c)] are arbitrary scaled for better comparison.
		The cyan scale bar in [(a)-(c)] is 200~$\mu$m.
		(d) Comparison of intensities projected along the vertical direction in [(a), blue circles], [(b), red circles] and [(c), black dots].
		(e) Comparison of intensities projected along the vertical direction obtained from the ghost imaging simulations with different speckle size (FWHM) [from 30~$\mu$m to 200~$\mu$m] compared to the projection of the near-field image of the sample collected by the bucket detector convoluted with one pixel point spread function [black line].
	}
	\label{fig:result}
\end{figure}

By implementing Eq.~\eqref{eq:GI} to 20,000 realizations of position III, we obtained the ghost image shown in Fig.~\ref{fig:result}(a).
Yet, this image hardly resembled the double bar shape illustrated in Fig.~\ref{fig:result}(c).
To improve the result, we further reduced the inhomogeneity of the background and increased the speckle contrast by means of an intensity normalization procedure [see Appendix C].
The result of this normalization, applied separately to each realization of the reference region, clearly demonstrates a more uniform and flat time-averaged background [see Fig.~\ref{fig:beam}(d)] with respect to the original one [see Fig.~\ref{fig:beam}(c)].
After applying Eq.~\eqref{eq:GI} and modified background displayed in Fig.~\ref{fig:beam}(d), we obtained the ghost image shown in Fig.~\ref{fig:result}(b).
The resemblance of the latter with the expected result [see Fig.~\ref{fig:result}(c)] demonstrates the importance of a proper consideration of the background.
To translate the qualitative visual assessment of Figs.~\ref{fig:result}(a,b) to a quantitative estimation, we projected the intensities along the vertical direction of the detector, obtaining a one-dimensional curve.
In Fig.~\ref{fig:result}(d) we compare the projections of the near field image of the object, measured using the bucket detector as a pixelized detector and shown in Fig.~\ref{fig:result}(c), and its ghost images.
While the curve obtained without intensity normalization [Fig.~\ref{fig:result}(a)] does not perfectly fit the bucket line, the one with normalization [Fig.~\ref{fig:result}(b)] nicely maps the expected result.

Apart from the inhomogeneity of the background, additional aspects contributed to the reduction of the maximum resolution achievable in our experiment.
First, speckle patterns corresponding to different TG orders were not confined enough in space and thus overlapped [see Figs.~\ref{fig:beam}(a,b)].
To overcome this obstacle and increase the resolution, an improved beam splitter with larger beam separation angle should be used in future experiments.
Second, the size of the speckles produced by the diffuser should ideally match the size of the detector pixel, which in turn should be as small as possible for higher resolution.
A third factor that influences the result is the number of realizations used to recover the ghost image.
In fact, to obtain a good signal-to-noise ratio, the number of realizations should be either bigger than the number of pixels in the reference area or the number of speckles probing the sample should exceed the number of pixels in the reference detector~\cite{Erkmen2009, Katz2009, Pelliccia2016}.

To assess the role of the speckle size on the resolution in GI, we performed simulations varying this parameter and using the wavelength, sample size, and geometry as in our experiment [see Fig.~\ref{fig:GI}(b)].
In particular, we generated 20,000 realizations of a random speckle pattern, with FWHM size of the speckles ranging from 30~$\mu$m to 200~$\mu$m at the detector, and propagated it to the reference and bucket detectors [see Appendix D].
As illustrated by Fig.~\ref{fig:result}(e), our simulations demonstrate a strong decrease in resolution when the speckle size approaches the size of the characteristic dimensions of the object, which in our case is 200 $\mu$m.
On the contrary, the resolution improves when the size of the speckles decreases.

In summary, we demonstrated ghost imaging at an XUV free-electron laser facility, employing a diffuser to generate a speckle pattern and a transmission grating to produce identical copies of the speckle beam.
We obtained the ghost image of a double slit structure by computing the correlations from 20,000 frames between a pixelized reference area and the integral of a bucket region.
We investigated the effect of an inhomogenous background on the reconstruction of the ghost image by probing different positions relative to the direct beam.
We finally demonstrated a successful reconstruction of the double slit structure for the position characterized by the most uniform background.
We also found, with the support of simulations, that a further improvement of the achievable resolution may be obtained by reducing the speckle size of the beam.

With this experiment, we pave the way for high-resolution ghost imaging at free electron laser facilities.
This technique, belonging to the realm of quantum based imaging protocols, is expected to become an important tool for imaging, in particular of biological samples, as the method allows a strong reduction of the intensity of the beam probing the sample.

\bigskip
\begin{acknowledgements}
We acknowledge E. Weckert for fruitful discussions and support of the project and M. Rose for careful reading of the manuscript.
We are grateful to the FLASH machine operators and the technical staff at FLASH for excellent FEL conditions.
This work was supported by the Helmholtz Associations Initiative and Networking Fund and the Russian Science Foundation (project No. 18-41-06001).
MB is funded through the Helmholtz Association via grant VH-NG-1105.
\end{acknowledgements}

\newpage
\appendix

\section{Ghost imaging reconstructions at positions I and II}
\begin{figure}[h!]
	\centering
	
	\includegraphics[scale = 0.52]{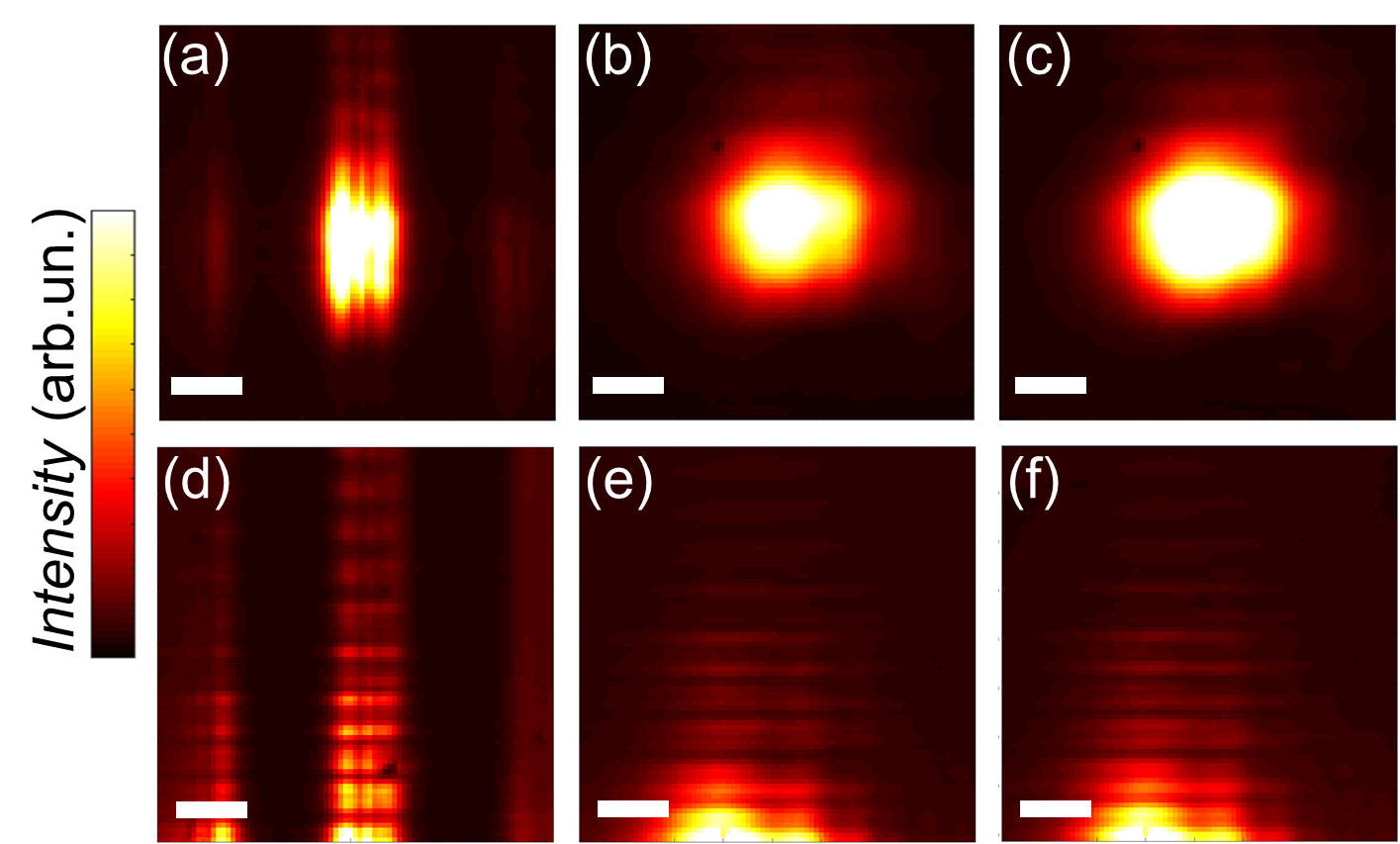}
	
	\caption{The results of GI from positions I and II after 20,000 realizations.
		(a-c) Position I. Time-averaged intensity of the bucket area (a) and the reference area (b). The result of ghost imaging (c).
		(d-f) Position II. Time-averaged intensity of the bucket area (d) and the reference area (e). The result of ghost imaging (f).
		The scale bar is 200 $\mu$m.}
	\label{fig:resI_II}
\end{figure}

As discussed in the main part of the paper, three regions, defined in Fig.~\ref{fig:beam}(b), were explored during our experiment.
While a satisfactory reconstruction was obtained for region III, no image resembling the sample was obtained for regions I and II, as illustrated by Fig.~\ref{fig:resI_II}.

%

\section{Correlation of bucket and reference beams}
For GI, identical speckle beams must illuminate the sample and the reference area at each pulse.
To verify this condition for our experiment we performed cross-correlation analysis of the measured speckle patterns.
The general definition of cross-correlation $\gamma(x, y)$ of an arbitrary template $B(x, y)$ of size $(X, Y)$ and a two-dimensional dataset $I(x, y)$ for a single realization can be expressed as
\begin{equation}
\label{eq:CC}
\gamma(x, y) = \sum_{x^\prime = 1}^{X} \sum_{y^\prime = 1}^{Y} I(x + x^\prime, y + y^\prime) \, B(x^\prime, y^\prime) \ .
\end{equation}
\begin{figure}[h!]
	\centering
	
	\includegraphics[width = 0.8\columnwidth]{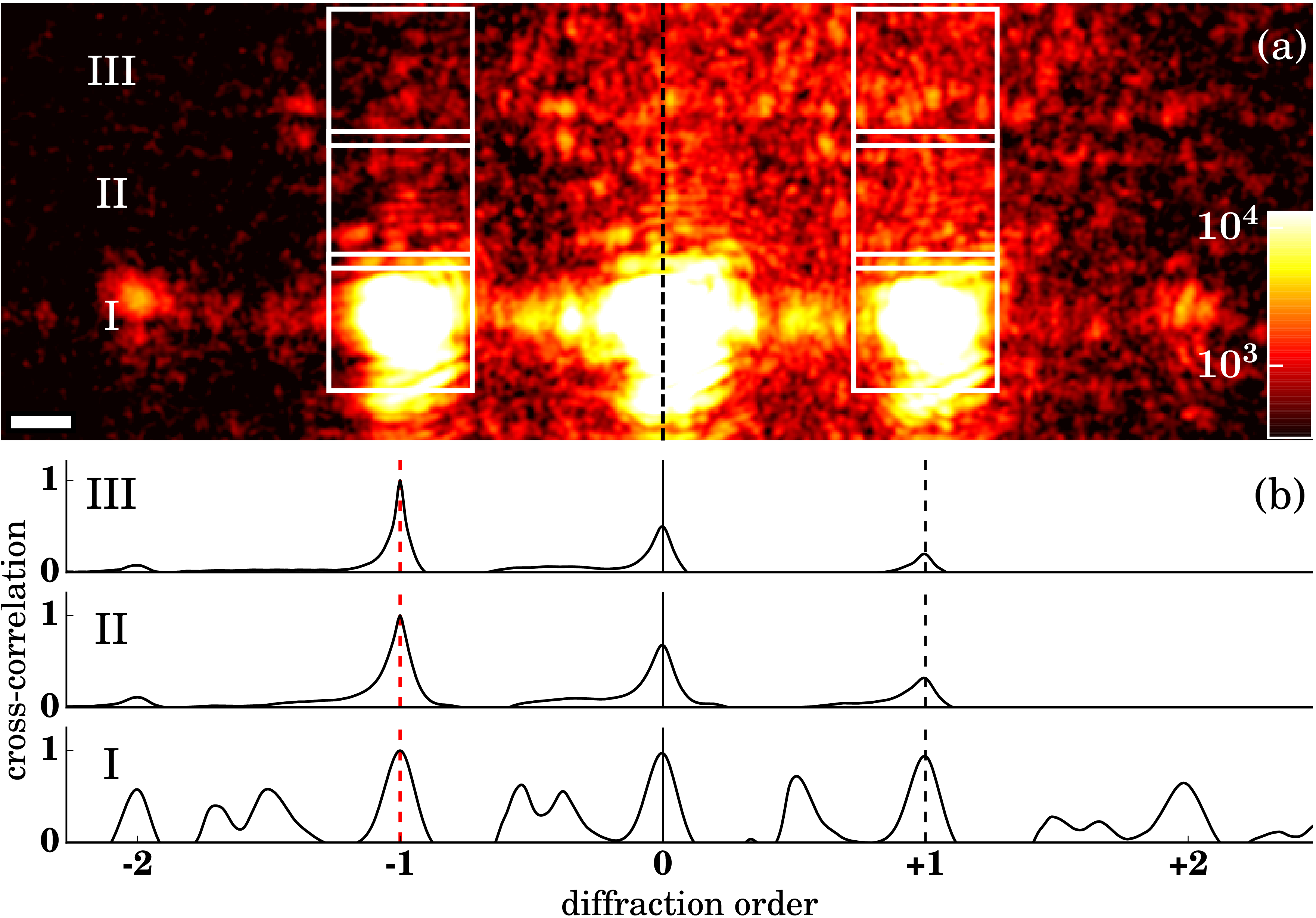}
	
	\caption{
		(a) Intensity at a selected position of the diffuser, shown to highlight the speckle pattern.
		White boxes define different probed positions, the colorbar defines the number of detected photons whereas the length of the white bar equal to 500 $\mu$m.
		(b) Results of the time-average zero mean normalized cross-correlation function, ranging from $+1$ [perfect correlation] to $0$ [no correlation].
		Each of the boxes corresponding to $-1$-order in (a) is correlated with every detector position and the line through each box center is drawn.
		The structure of the diffracted beam is especially evident from position I, demonstrating strong correlation for $\pm 1$, $\pm 2$, and $0$-order as well as half-orders [diffraction from the second harmonic in the FEL beam].
	}
	\label{fig:beamS}
\end{figure}

\noindent
To account for different illumination conditions, e.g. strong beam gradients, $\gamma(x, y)$ can be zero-normalized [i.e. zero mean value] as~\cite{Bernstein1983}
\begin{equation}
\label{eq:ZNCC}
\gamma^\prime(x, y) = \sum_{x^\prime = 1}^{X} \sum_{y^\prime = 1}^{Y} \dfrac{I(x + x^\prime, y + y^\prime) - \mu_I({x, y})}{\sigma_I({x, y})} \, \dfrac{B(x^\prime, y^\prime) - \mu_B}{\sigma_B} \ ,
\end{equation}
where $\mu_B$ is the mean and $\sigma_B$ is the standard deviation of the template $B(x, y)$.
The local mean $\mu_I$ and the local standard deviation $\sigma_I$ of the dataset $I(x, y)$, calculated over the template region $B(x, y)$, are defined as
\begin{equation*}
\begin{split}
&\mu_I({x, y}) = \dfrac{1}{X Y} \sum_{x^\prime = 1}^{X} \sum_{y^\prime = 1}^{Y} I(x + x^\prime, y + y^\prime)\ ,\\[2ex]
&\sigma_I({x, y}) = \dfrac{1}{X Y} \sqrt{\sum_{x^\prime = 1}^{X} \sum_{y^\prime = 1}^{Y} \Big[I(x + x^\prime, y + y^\prime) - \mu_I({x, y})\Big]^2} \ .
\end{split}
\end{equation*}
The zero-normalized cross-correlation function is defined in the range [$-1$, $+1$], where $-1$ implies anti-correlation, $0$ no correlation and $+1$ full correlation.

To quantify the correlation among different positions on the detector [regions I, II, and III shown in  Fig.~\ref{fig:beamS}(a)] we implemented the zero mean normalized cross-correlation of the bucket region intensity $B(x,y)$ at the $-1$-order, where the sample was positioned, and detected intensity $I(x,y)$.
In particular, Fig.~\ref{fig:beamS}(b) shows the average zero mean normalized cross-correlation over 2,115 realizations.

The strongest correlation is between the $-1$-order, illuminating the sample, and the $0$-order, which was blocked by the sample holder during the experiment to avoid detector saturation.
The correlation is also noticeable between the $-1$-order and $+1$-order, satisfying therefore the necessary condition to perform ghost imaging.

\section{Intensity Normalization}
To improve our ghost image -- in particular to decrease the inhomogeneity of the background and enhancing the contrast of the speckles -- we normalized intensities in each realization of the reference area by implementing the \textit{additive pattern method} developed in~\cite{IntensityNormal_Main}.
Within this formalism, the image is described as
\begin{equation}
I(x,y) = \alpha(x,y) +  \beta(x,y) I^{corr}(x,y) \ ,
\end{equation}
where $\alpha(x,y)$ and $\beta(x,y)$ are the background and modulation of light respectively and $I^{corr}(x,y)$ is the background-corrected intensity used for further analysis.

\begin{figure}[h!]
	\centering
	
	\includegraphics[scale = 0.6]{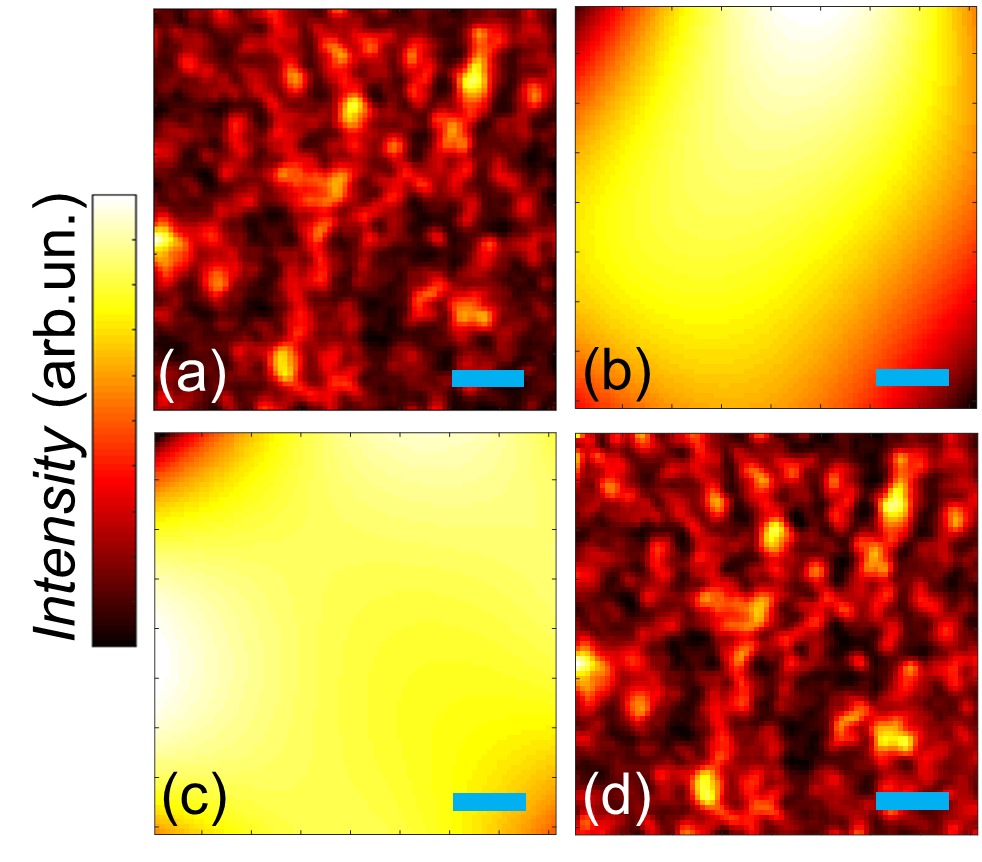}
	\caption{The intensity normalization procedure in the reference area corresponding to position III.
		(a) A single realization as measured;
		(b) the background $\alpha(x,y)$ of (a), extracted by polynomial fitting of (a);
		(c) the light modulation $\beta(x,y)$;
		(d) the final result of the procedure, function $I^{corr}(x,y)$.
		The cyan scale bar is 200 $\mu$m.
	}
	\label{fig:INproc}
\end{figure}

Elements described above are reported in Fig.~\ref{fig:INproc} [the detailed procedure is described in subsection 2.1 of~\cite{IntensityNormal_Main}].
In particular, a single realization $I(x,y)$ of the reference detector is shown in Fig.~\ref{fig:INproc}(a).
From a polynomial fitting of $I(x,y)$, the background $\alpha(x,y)$ [Fig.~\ref{fig:INproc}(b)] as well as the light modulation $\beta(x,y)$ [Fig.~\ref{fig:INproc}(c)] are extracted.
Finally, the result of the procedure, the function $I^{corr}(x,y)$ which is then used for GI reconstructions instead of $I(x,y)$, is illustrated in Fig.~\ref{fig:INproc}(d).

\section{Ghost imaging simulations}
\begin{figure}[h!]
	\centering
	
	\includegraphics[scale = 0.6]{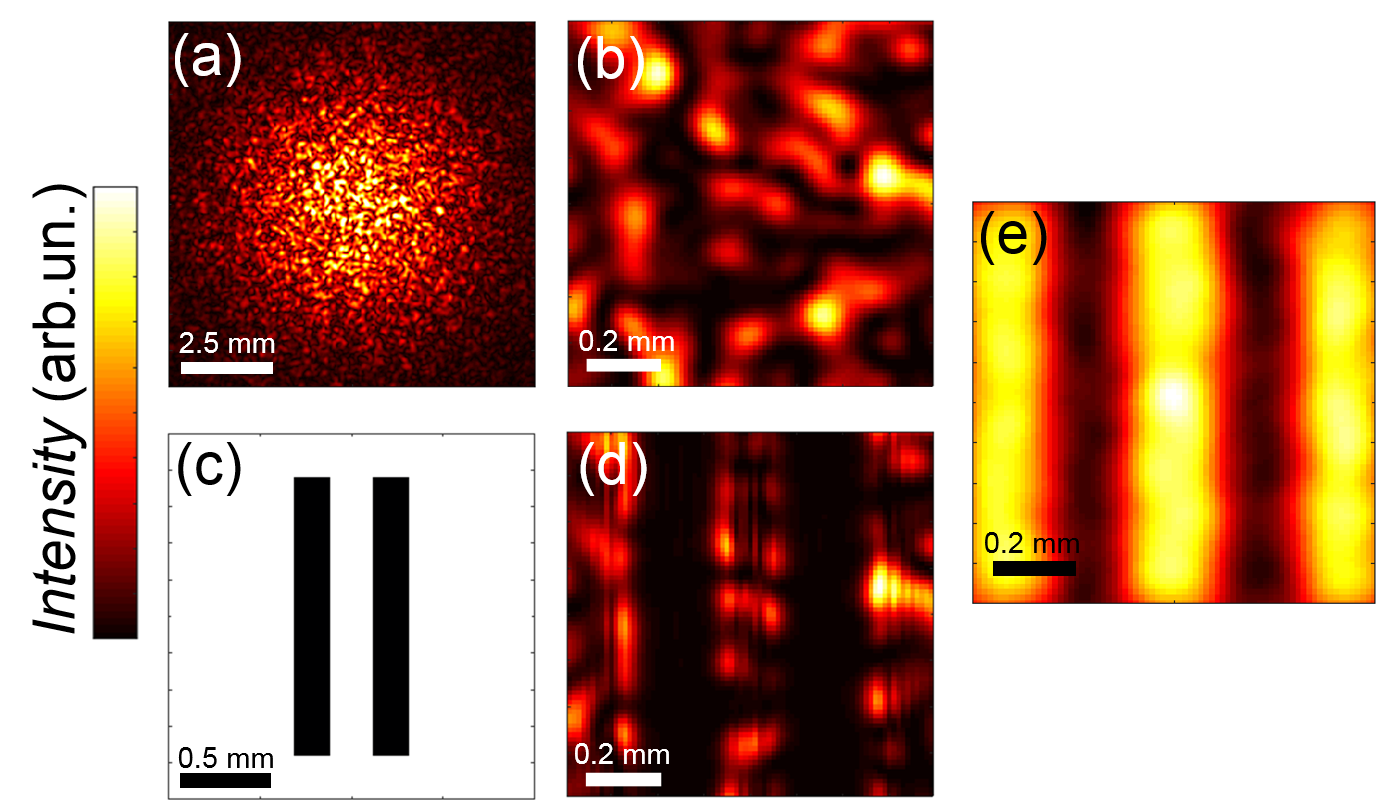}
	\caption{Simulation procedure. (a) An initial beam consisting of a random speckle patterns; (b) the same propagated to the reference detector; (c) the simulated sample [fully opaque, with the same geometry as the one employed in our experiment]; (d) the beam defined in (a) propagated though the sample; (e) the result of ghost imaging reconstruction when using a speckle size of 100 $\mu$m FWHM.}
	\label{fig:SimulBeam}
\end{figure}

Ghost imaging simulations are performed using the same wavelength, sample size, and geometry described in the main part of the paper.
However, to avoid background inhomogeneities, the full-width at half-maximum (FWHM) of the beam impinging on the diffuser is considered to be 40 times larger than the experimental one.\\
To generate a spatially incoherent beam, emulating the one produced by the diffuser, we follow the procedure described in ~\cite{Pfeifer2010}.
In particular, we initially generate an electric field having Gaussian amplitude ${A_{f}}(x,y)$ of 3.5 mm FWHM and random phase field ${\varphi_{f}}(x,y) \in [-\pi,\pi]$,
\begin{equation}
\label{eq:Beam1}
{E_{f}}(x,y) = {A_{f}}(x,y)\exp(i{\varphi_{f}}(x,y)) \ .
\end{equation}
We then Fourier transform this field to frequency space ($f_{X}, f_{Y}$)
\begin{equation}
\label{eq:Beam2}
E_{0}(f_{X},f_{Y}) = \mathcal{F}\left[{E_{f}}(x,y)\right]
\end{equation}
and multiply the phase by a Gaussian filter $F_{0}(f_{X},f_{Y})$
\begin{equation}
E_{f}(f_{X},f_{Y}) = F_{0}(f_{X},f_{Y})\frac{E_{0}(f_{X},f_{Y})}{\vert E_{0}(f_{X},f_{Y}) \vert} \ .
\end{equation}
The final speckle pattern is obtained as the inverse Fourier transform
\begin{equation}
{E_{f}}(x,y) = \mathcal{F}^{-1}\left[E_{f}(f_{X},f_{Y})\right] \ .
\end{equation}
By this procedure, the variance of the average size of speckles at diffuser position is the inverse of the variance of the Gaussian filter $F_{0}(f_{X},f_{Y})$ [see an exemplificative realization of the speckle pattern on the diffuser in Fig.~\ref{fig:SimulBeam}(a)].

This speckled field ${E_{f}}(x,y)$ is then propagated using the angular spectrum method~\cite{Goodman1996IFO, Matsushima}.
First, the angular spectrum $A_{0}(f_{X},f_{Y},0)$ is generated from the field $E_{f}(x,y,0)$,
\begin{equation}
A_{0}(f_{X},f_{Y},0)= \iint E_{f}(x,y,0)\mathrm{exp}\left(-i2\pi (f_{X}x+f_{Y}y)\right)\,\mathrm{d}x\,\mathrm{d}y \ .
\end{equation}
Then, to determine $E_{z}(x,y,z)$ at the distance $z$ from the diffuser ($z = 0$), the following equations are considered
\begin{equation}
\begin{split}
A_{z}(f_{X},f_{Y},z) &= A_{0}(f_{X},f_{Y},0)\exp\left(i(\frac{2\pi}{\lambda})z\sqrt{1-(\lambda f_{X})^{2}-(\lambda f_{Y})^{2}}\right) \ ,\\
E_{z}(x,y,z) &= \iint A_{z}(f_{X},f_{Y},z)\exp\left(i2\pi (f_{X}x+f_{Y}y)\right)\,\mathrm{d}f_{X}\,\mathrm{d}f_{Y} \ .
\end{split}
\end{equation}
According to our experimental geometry, we propagate the reference beam $E_{z}(x,y,z)$ from the diffuser to the detector, for $z$ = 1.79 m [see Fig.~\ref{fig:SimulBeam}(b)].

To account for the presence of the sample, we first propagate the beam from the diffuser to the object [$E_{z_{1}}$($x,y,z_{1}$ = 1.49 m)] and then we multiply this amplitude by the sample function $O(x,y)$ [illustrated in Fig.~\ref{fig:SimulBeam}(c)].
The result is further propagated to the detector $E_{z_{2}}$($x,y,z_{2}$ = 0.3 $\mathrm{m}$) [Fig.~\ref{fig:SimulBeam}(d)].
Finally, ghost imaging calculations are performed on 20,000 realizations generated using the procedure described above.
The result of the reconstruction with a speckle size [at the detector position] of 100 $\mu$m FWHM -- corresponding to experimental values -- is shown in Fig.~\ref{fig:SimulBeam}(e). The whole procedure is implemented in the MATLAB package.

\newpage

\bibliography{reference}

\end{document}